# Flexible transistors exploiting P3HT on paper substrates and graphene oxide films as gate dielectrics: proof of concept

Luca Valentini[1*], Marta Cardinali,[1] Mirjana Mladjenovic,[2] Petar Uskokovic,[2] Federico Alimenti,[3] Luca Roselli,[3] and Josè Kenny[1]

[1]Università di Perugia, Dipartimento di Ingegneria Civile e Ambientale, Strada di Pentima 4, INSTM, UdR Perugia, 05100 Terni Italy
[2]Faculty of Technology and Metallurgy University of Belgrade Karnegijeva 4 11000 Belgrade Serbia
[3]Università di Perugia, Dipartimento di Ingegneria Elettronica e dell'Informazione, Via G. Duranti 93, 06123 Perugia Italy

**In this paper we report: the use of GO in aqueous solution as a gate dielectric material, its application to a MOS transistor based on organic semiconductor and the use of paper as substrate material.**

Flexible electronics, a technology that has witnessed significant attention and investment in diverse engineering research and development fields, is increasingly important in today's growing market for devices demanding shape adaptability, light weight and space savings.[1-4] Paper, an organic-based substrate, is universally available: the high demand and the mass production of paper has made it one of the cheapest materials.[1] The possibility to integrate electronic and optoelectronic functions within the production methods of the paper industry is of current interest to enhance and to add new functionalities to paper.[2-4]

Graphene oxide[5-7] (GO) is a water soluble insulating material that can be obtained by oxidizing graphite. Few-layer GO films can be deposited and bonded to various substrates by water solution-based processing at low temperatures, which will meet the technological challenges for delivering low-cost and thin dielectric material layers. Electrical characteristics of graphite oxide thin films deposited on a p-type silicon substrate were investigated to explore its potential application as a dielectric material in field-effect transistors.[8] More recently high-performance graphene-based thin film transistor was fabricated on plastic substrates using graphene as active layer, graphene oxide as dielectric and graphene electrodes.[9] Furthermore, the high cost of the single crystal substrates and the ultra high vacuum conditions necessary for graphene growth as well as the graphene transferring limit the use of these methods for fast, cheap and large scale applications.

Poly(3-hexylthiophene) (P3HT) is a semiconducting polymer with extensive use in organic electronics,[10,11] it has a regular end-to-end arrangement of side chain allowing efficient π-π stacking of the conjugated backbones with the hole mobility in the range of $10^{-3}$-$10^{-1}$ $cm^2V^{-1}s^{-1}$. Poly(alkylthiophenes) play a major role in the development of polymer field effect transistors.[12,13] Poly(3-hexylthiophene) is hydrophobic at neutral state, which is attributed to the alkyl side group.[14] The challenge in using a layer sequence mainly concerns the drop-casting of soluble dielectric materials on top of the hydrophobic polymer semiconducting layer.

In the assembly of an organic thin film transistor, an insulating film is often deposited on top of the semiconductor layer. This insulating layer can have packaging and gas barrier effects on semiconductor films that strongly interact with environmental species such as oxygen or moisture, which is reflected in large variations of the electrical properties exhibited by the device.[15,16] Additionally, the insulating layer on top of the thin film transistor structure can work as an effective mechanical and chemical protection of the devices from subsequent processes.[17]

The main focus of this work is the localization of GO solution on the hydrophobic layer made of regioregular poly(3-hexylthiophene). At our best of knowledge the deposition from water-based solutions of GO as insulating top layer for organic electronics is not yet realized. With respect to the $SiO_2$ that at our best knowledge has never been deposited on paper substrates, this method is a viable solution toward paper based organic transistors. Our transistor is completely planar, i.e. it does not use any backgate structure. This geometry is different with respect to that of most thin film transistors already published[18] and has been selected to facilitate the device connection to the external circuitry. So it is a fair question to ask whether the use of GO thin films act as gate dielectrics for planar top-gate field-effect transistors via solution-based deposition on paper in ambient conditions.

Here, in analogy to the widespread use of silicon dioxide in silicon microelectronic industry, we report the proof-of-concept for the use of graphene oxide as a gate dielectrics for polymer field-effect transistor via a fast and simple solution-based deposition on paper.

Graphene oxide were purchased from Cheaptubes (thickness 1-5 nm estimated by AFM). Water dispersion (1mg/mL) was prepared and tip sonicated (750W, 60% amplitude) for 1 hr to yield a yellow suspension. After this mixing process, the solution was transferred to a vial and it was centrifuged for 30 min at 9000 rpm. The supernatant of the dispersion was carefully extracted and separated from the residual visible at the bottom of the vial. This procedure was adopted because during reaction and processing, the graphene sheets are not only derivatized with oxygen-containing groups but also disaggregated into smaller pieces. As a result, the lateral sizes of the as-synthesized GO sheets are usually very polydisperse, ranging from a few nanometers to tens of micrometers, which may even vary from synthesis to synthesis. The centrifugation helps to remove such colloidal particles. The GO layer was annealed in vacuum for 1 hr at 100°C in order to remove water moisture.

Regioregular poly(3-hexylthiophene) (P3HT catalog number 698997 electronic grade, purchased from Aldrich, average molecular weight $M_n$ 54000-75000, >98% head-to-tail regioregular) was adopted for this research.

The semiconducting polymer layer was prepared by drop casting from chloroform solution with a concentration of 1mg/mL onto the paper substrate. This step was performed in a glove box with nitrogen streaming. The coating process is done at ambient temperature and P3HT films with different thicknesses between 1 um and 10 um for multiple drop casting trials were obtained. Then the source and drain electrodes, which are made up of Au, were deposited on top of P3HT film through a Teflon mask by thermal evaporation ($\approx 10^{-6}$ Torr) with an optimized thickness of 60 nm. The source and drain electrodes have a dimension of 8mm X 2.5mm.

The gate dielectric was obtained by depositing 1 mg/mL aqueous solution of graphene oxide onto P3HT coating paper using drop-and dry method. Large drops (i. e. $\approx$ 2 mm) of 2 μL aqueous solution were obtained. The thickness of the GO multilayered film was estimated to be about 200 nm by scanning electron microscopy with a lateral size of about 6mm. Thermally evaporated top gate contact (i. e. Au) has the dimension of 1mm X 1mm. Co-planar top-gate transistors were fabricated on multiple samples by varying the distance between source and drain electrodes from 6 mm down to 1 mm. A completed FET device with its schematic representation is shown in Figure 1.

The morphologies of the prepared samples were investigated by atomic force microscopy (AFM). AFM images were obtained in tapping mode. Contact angles were measured with an optical contact angle meter at room temperature. Water droplets were dropped carefully onto the surfaces and the contact angle was monitored. Electrical characteristics of the device were measured using Keithley 4200-SCS semiconductor characterization system.

Commercial photopaper (Mitsubishi – Electric, 250 μm thick, $\varepsilon_r \approx 3.2$) without any kind of surface treatment was used as substrate. We deposited P3HT on paper as reported in Figure 1 and contact it with gold source and drain pads. We investigated how processing parameters such as depositing multiple drops of graphene oxide affects the GO uniformity for the top gate dielectric layer.

Figure 2a shows the atomic force microscopy image of the paper substrate with P3HT coating layer. As expected from the observation through the optical microscope of bare paper substrate, the surface characteristic of the P3HT is very poor and the root-mean-square roughness of the P3HT surface is more than 31 nm. The AFM image of a GO-coated P3HT is displayed in Figure 2b. Compared to the P3HT surface, the surface of GO layer is relatively smooth with the root-mean-square roughness less than 11 nm. Graphene oxide film shows a pinhole-free structures with curvatures.

When a drop of a solution containing dispersed particles evaporates on a surface, it commonly leaves a dense, ring-like deposit along its perimeter.[19] This is the so-called "coffee ring effect", i.e., a distortion of the drops during solvent drying due to the interplay of ink viscosity and solute transport via solvent motion.[20] This is one of the most important phenomena affecting the homogeneity of dried solution drops. In order to prevent this, it is necessary to "freeze" the drops' geometry after they form a homogeneous and continuous film on the substrate.

Considering the structure of the P3HT it is expected that the modification of the hydrophilic paper surface with the P3HT alkyl chains can switch from hydrophilic to hydrophobic the behavior of the paper substrate as reported in Figures 2c and 2d. The lower wettability of the P3HT coated paper makes possible the localization of the drop of the GO solution avoiding its spanning. Dried GO drops with their carboxylic groups lead to a further switch of the surface wettability (Figure 2e). Inkjet printing is one of the most promising techniques for large-area fabrication of flexible plastic electronics. A key property of inks viable for printing is their ability to generate droplets. So the drop casting method simulates reasonably the inkjet mechanism. On the other hand ink viscosity, surface tension, density and nozzle diameter influence the spreading of the resulting liquid drops.[21] During printing, the primary drop may be followed by secondary (satellite) droplets. This needs to be avoided in drop-on-demand printing. We are working in our lab to control such physical parameters to obtain uniform droplets during the deposition.

We have also evaluated the out-of-plane electric properties of uniform drop cast GO film as shown in Figure 2f. For such characterization we applied a bias to the top electrode with the bottom electrode grounded. Figure 3a shows the bias dependent leakage current curves of such device. The electrical behavior in air indicates that the dielectric breakdown of the GO film in the out-of-plane direction occurs at a bias field larger than 1 MVcm$^{-1}$. Such dielectric breakdown field is comparable with that of oxide dielectrics.

Figure 3b shows the drain-source current ($I_{DS}$) versus drain-source voltage ($V_{DS}$) output characteristics of the transistors at various top-gate voltages ($V_{GS}$) as well as the transfer curve (Figure 3c). The electrical transport studies of the top-gated graphene transistors were carried out at room temperature.

Experiments with the source-drain distance of 6 mm, 4 mm and 2 mm, respectively showed a quasi linear behaviour of the drain-source current (see Supporting Information). The transistor behavior was not apparent in these cases because the channel is too resistive and the gate dielectric has only a minor control over the channel. Lowering the source-drain distance up to 1 mm, the device showed a field-effect response (Figure 3b). Our curves are in agreement with those observed by E. M. C. Fortunato et al.[22] where p-channel transistors layered on paper operate in depletion mode and do not exhibit hard saturation behavior. The reported field-effect mobility, µ, was extracted from the transfer curve in the saturation regime, i.e. at $V_{DS}$ = -25 V.

In particular the $I_{DS}$ of the film increased with $V_G$ approaching more negative values, thus suggesting hole conduction. The relationship between the molecular ordering of polymer layer and the electrical characteristics of the device has been extensively investigated.[21] Especially using P3HT as the active layer, there is a direct correlation between the molecular structure and the field effect mobility.[23] Therefore, the improvement of surface characteristics of the bare paper substrate should be considered to achieve high field effect mobility. In any case in our case the on/off ratio is larger than 10. Estimating a capacitance for the GO layer[9] of 25 nF/cm$^2$ and a ratio between the channel width and source/drain distance of 2.5, the saturation mobility, µ$_{sat}$, of the single component amounts to 0.043 cm$^2$V$^{-1}$s$^{-1}$for P3HT, which is in the range of mobility values reported for P3HT with high regioregularity by Sirringhaus et al.[23]

A new fabrication methodology for gate dielectrics printed on paper substrates has been developed. This novel methodology deals with the type of both substrate material and gate material. In particular, the paper substrate with its low cost and large amount is suitable for fast printing processes such as direct write methodologies or inkjet printing of electronics. Graphene oxide based solutions can be used efficiently in conjunction with appropriate surface energy of semiconducting polymers to produce top gate dielectric multilayer electronics on paper. Such a design methodology has been demonstrated by a measured prototype.

1 A. C. Siegel, S. T. Phillips, M. D. Dickey, N. Lu, Z. Suo, G. M. Whitesides, *Adv. Funct. Mater.* 2010, **20**, 28.

2 L. Yang, A. Rida, R. Vyas, M. M. Tentzeris, *IEEE Trans. Microw. Theory Tech.* 2007, **55**, 2894.

3 E. Fortunato, N. Correia, P. Barquinha, L. Pereira, G. Goncalves, R. Martins, *IEEE Electron Device Lett.* 2008, **29**, 988.

4 S. Couderc, O. Ducloux, B. J. Kim, T. Someya, *J. Micromech. Microeng.* 2009, **19**, 1.

5 L. J. Cote, R. Cruz-Silva and J. Huang, *J. Am. Chem. Soc.* 2009, **131**, 11027.

6 K.S. Subrahmanyam, P. Kumar, A. Nag and C.N.R. Rao, *Solid State Commun.* 2010, **150**, 1774.

7 K. A. Mkhoyan, A. W. Contryman, J. Silcox, D. A. Stewart, G. Eda, C. Mattevi, S. Miller, M. Chhowalla, *Nano Lett.* 2009, **9**, 1058.

8 I-Y Lee, E. S. Kannan, G.-H. Kim *Appl. Phys. Lett.* 2009, **95**, 263308.

9 S.-K. Lee, H. Y. Jang, S. Jang, E. Choi, B. H. Hong, J. Lee, S. Park, J.-H. Ahn *Nano Lett.* 2012, **12**, 3472.

10 G. Li, V. Shrotriya, J. Huang, Y. Yao, T. Moriarty, K. Emery, Y. Yang, *Nat. Mater.* 2005, **4**, 864.

11 F. Padinger, R. S. Rittberger, N. S. Sariciftci, *Adv. Func. Mater.* 2003, **13**, 85.

12 A. C. Arias, S. E. Ready, R. Lujan, W. S. Wong, K. E. Paul, A. Salleo, M. L. Chabinyc, R. Apte, R. A. Street, Y. Wu, P. Liu, B. Ong, *Appl. Phys. Lett.* 2004, **85**, 3304.

13 M. L. Chabinyc, W. S. Wong, A. C. Arias, S. Ready, R. A. Lujan, J. H. Daniel, B. Krusor, R. B. Apte, A. Salleo, R. A. Street, *Proc. IEEE* 2005, **93**, 1491.

14 X. Wang, T. Ederth, O. Inganäs, *Langmuir* 2006, **22**, 9287

15 C. R. Kagan, P. Andry, Thin-film transistors, Marcel Dekker, Inc., New York 2003.

16 M. Kim, J. H. Jeong, H. J. Lee, T. K. Ahn, H. S. Shin, J. S. Park, J. K. Jeong, Y. G. Mo, H. D. Kim, *Appl. Phys. Lett.* 2007, **90**, 2114.

17 A. Sato, K. Abe, R. Hayashi, H. Kumomi, K. Nomura, T. Kamiya, M. Hirano, H. Hosono, *Appl. Phys. Lett.* 2009, **94**, 133502.

18 E. Fortunato, P. Barquinha, R. Martins *Adv. Mater.* 2012, **24**, 2945.

19 M. Singh, H. M. Haverinen, P. Dhagat, G. E. Jabbour, *Adv. Mater.* 2010, **22**, 673.

20 R. D. Deegan, O. Bakajin, T. F. Dupont, G. Huber, S. R. Nagel, T. A. Witten, *Nature* 1997, **389**, 827.

21 B. Derby, N. Reis, *MRS Bull.* 2003, **28**, 815.

22 R. F. P. Martins, A. Ahnood, N. Correia, L. M. N. P. Pereira, R. Barros, P. M. C. B. Barquinha, R. Costa, I. M. M. Ferreira, A. Nathan, E. E. M. C. Fortunato, *Adv. Funct. Mater.* 2012 DOI: 10.1002/adfm.201202907

23 H. Sirringhaus, P. J. Brown, R. H. Friend, M. M. Nielsen, K. Bechgaard, B. M. W. Langeveld-Voss, A. J. H. Spiering, R. A. J. Janssen, E. W. Meijer, *Synth. Met.* 2000, **111**, 129.

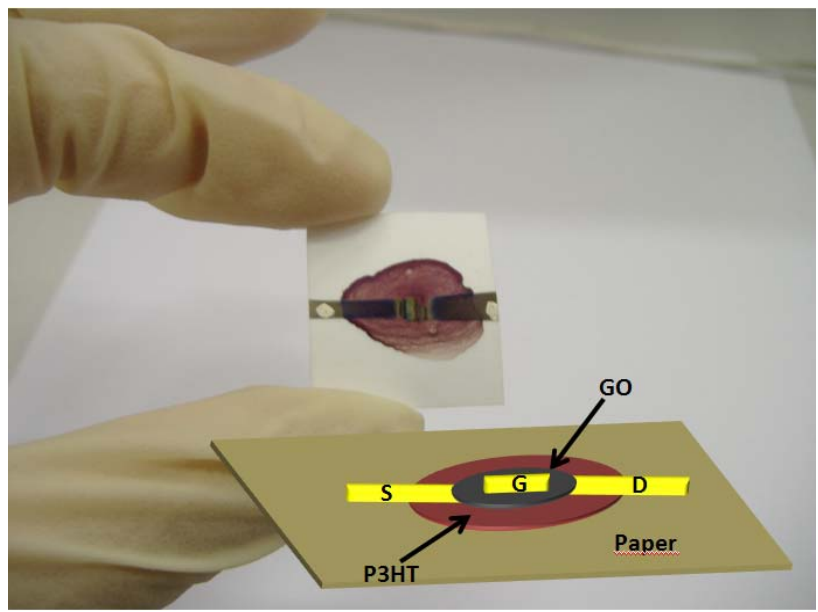
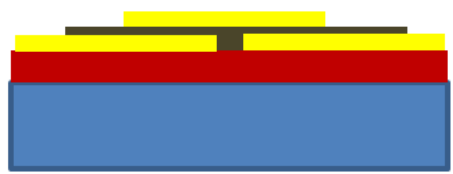

Figure 1. Schematic diagram of completed device; co-planar top gate structure adopted for our device with the representation of the substrate (blue), electrodes (yellow), semiconductor (brown) and gate dielectric layer (grey).

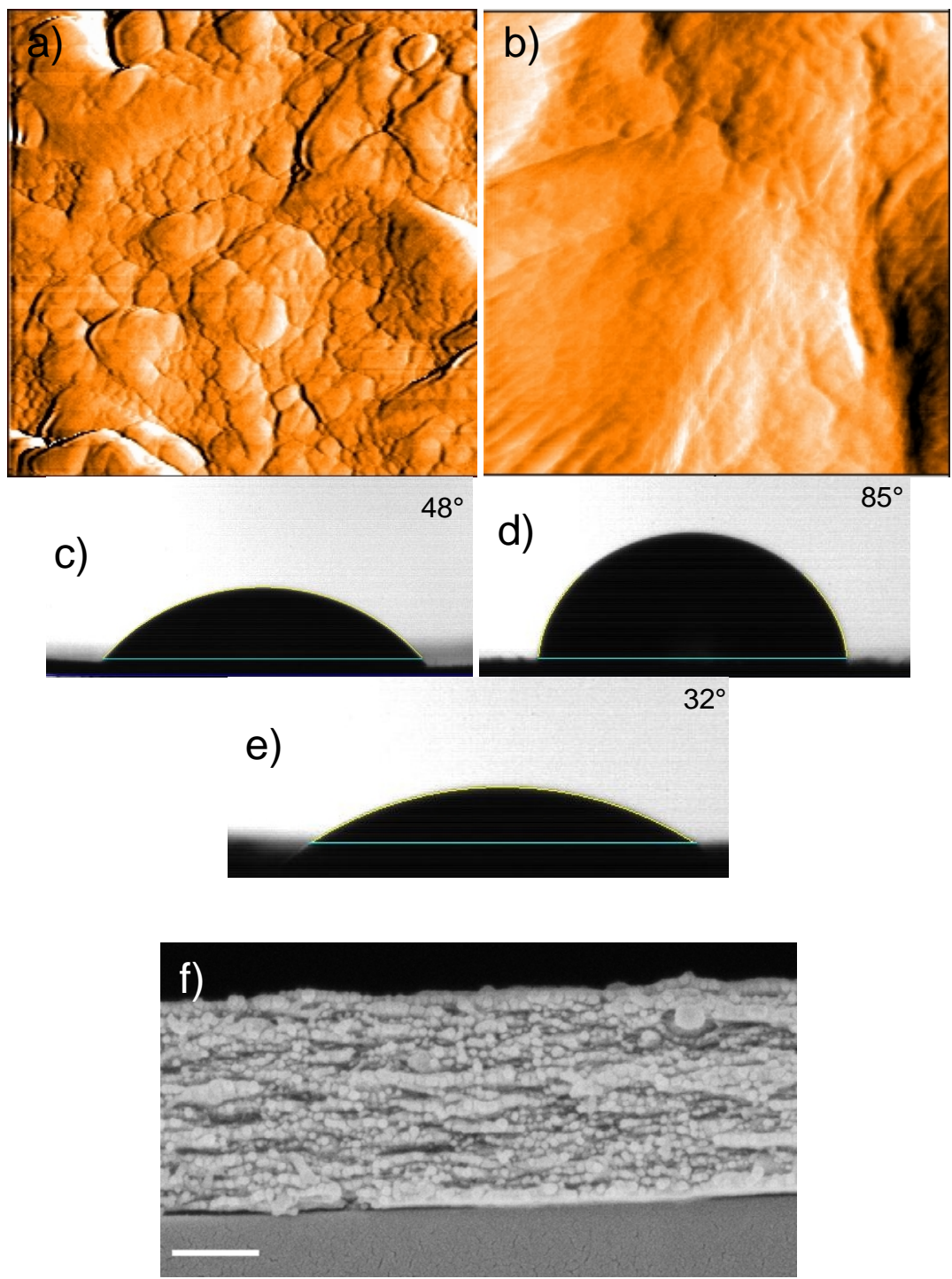

Figure 2. The AFM images (35 μm X 35 μm) of (a) P3HT deposited on paper (thickness of about 6 μm) and (b) multi-layer GO film deposited on the P3HT coated paper. Contact angle images of (c) neat paper substrate, (d) P3HT deposited on paper and (e) multi-layer GO film deposited on the P3HT coated paper. The contact angle values are reported on the right-hand side of the images. (f) Cross-section FESEM image of the fabricated 200 nm thick GO dielectric layer. The bright dots are due to the Au metallization for the visualization of the GO insulating layer. The scale bar indicates 100 nm.

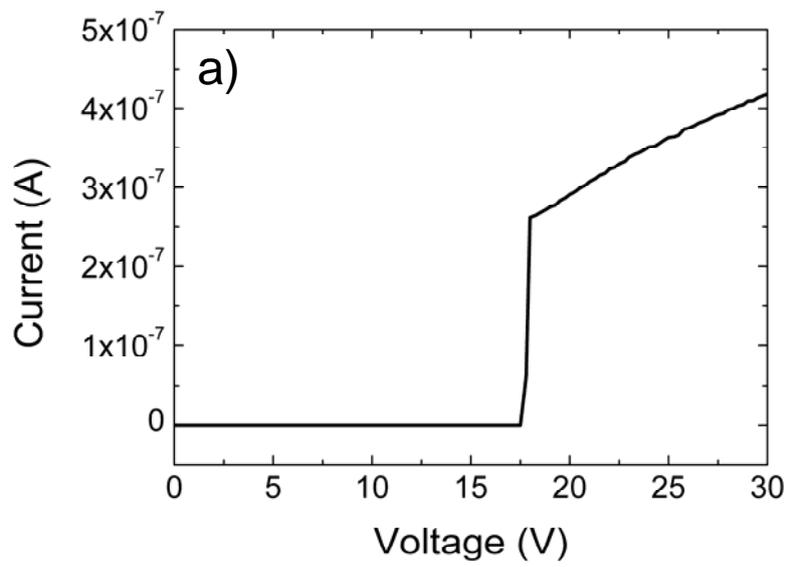

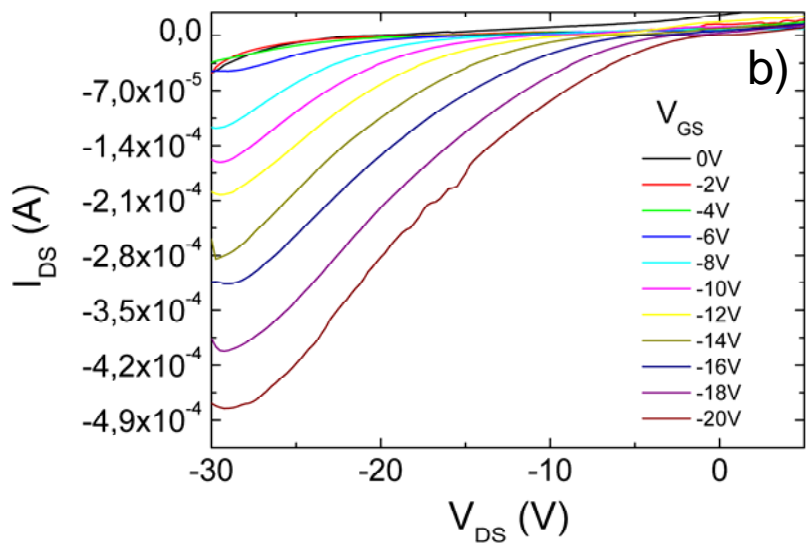

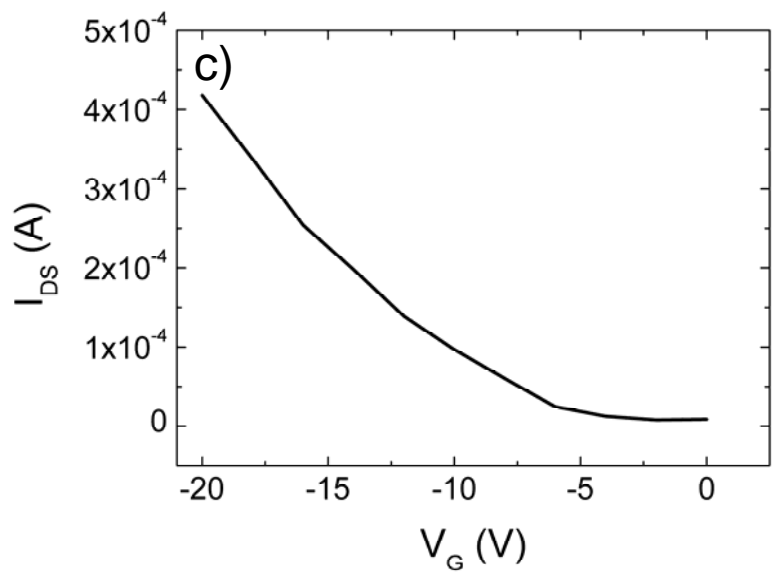

Figure 3. (a) Bias dependent leakage current (I-V) curve of GO film measured in air. (b) Output ($I_{DS}$–$V_{DS}$) and (c) transfer ($I_{DS}$–$V_{GS}$) curves of the device with graphene oxide film as gate dielectric (channel width=2.5 mm, source-drain distance=1 mm) having P3HT as the semiconductor.